\documentclass[11pt]{JHEP3} 

\usepackage{latexsym}
\usepackage{epsfig}
\usepackage{amssymb}
\usepackage{graphicx}

\newcommand{\lp}{\left(}
\newcommand{\rp}{\right)}
\newcommand{\lb}{\left[}
\newcommand{\rb}{\right]}

\newcommand{\ba}{\begin{eqnarray}}
\newcommand{\ea}{\end{eqnarray}}
\newcommand{\be}{\begin{equation}}
\newcommand{\ee}{\end{equation}}

\newcommand{\al}{\alpha}
\newcommand{\bt}{\beta}

\newcommand{\sa}{\sigma}

\newcommand{\La}{\Lambda}

\title{Constraining Entropic Cosmology}

\author{Tomi S. Koivisto \\ 
Institute for Theoretical Physics and the Spinoza Institute, Utrecht University, \\
Leuvenlaan 4, Postbus 80.195, 3508 TD Utrecht, The Netherlands \\
E-mail: \email{t.s.koivisto@uu.nl}}
\author{David F. Mota \\ 
Institute of Theoretical Astrophysics, University of Oslo, 0315 Oslo, Norway \\
E-mail: \email{d.f.mota@astro.uio.no}}
\author{Miguel Zumalac\'arregui \\ 
Institute of Cosmos Sciences (ICC-IEEC), University of Barcelona \\ Marti i Franques 1, E-08028 Barcelona, Spain \\
E-mail: \email{miguelzuma@icc.ub.edu}}


\preprint{1011.2226}

\abstract{
It has been recently proposed that the interpretation of gravity as an emergent, entropic phenomenon might have nontrivial implications to cosmology. Here several such approaches are investigated and the underlying assumptions that must be made in order to constrain them by the BBN, SneIa, BAO and CMB data are clarified. Present models of inflation or dark energy are ruled out by the data. Constraints are derived on phenomenological parameterizations of modified Friedmann equations and some features of entropic scenarios regarding the growth of perturbations, the no-go theorem for entropic inflation and the possible violation of the Bekenstein bound for the entropy of the Universe are discussed and clarified.
}

\begin{document}

\section{Introduction}

The notion of gravity as an emergent force has been contemplated for a long time \cite{Visser:2002ew}. The
derivation of gravitational field equations from thermodynamics by reference \cite{Jacobson:1995ab} supports this, and has lead to further substantial hints of evidence for the idea \cite{Padmanabhan:2009vy}.
Recently, the proposal was put forward that gravity is a thermodynamic phenomenon
emerging from the holographic principle \cite{Verlinde:2010hp}. It was argued that the Newton's law of gravitation can be understood as
an entropic force caused by the change of information holographically stored on a screen when material bodies are
moving with respect to the screen. This is described by the first law of thermodynamics, $F\Delta x = T\Delta S$,
connecting the force $F$ and the displacement $\Delta x$ to the temperature $T$ of the screen and the change of
its entropy, $\Delta S$. $T$ can be then identified with the Unruh temperature without referring
to a horizon. Postulating $\Delta S= 2\pi m\Delta x$, $m$ being a particle mass, Newton's second law follows.
More to the point, assuming equipartition of the energy \cite{Padmanabhan:2009kr} given by the enclosed mass,
Newtonian gravitation emerges. See reference \cite{Makea:2010zt} for subtly different viewpoints.
It has also been argued that the entropic scenario fails to reproduce the quantum states of gravitationally trapped neutrons \cite{Kobakhidze:2010mn}, which experimentally match the predictions of Newtonian gravity. If this is confirmed the entropic scenario would be ruled out.

Cosmology has been also considered in this framework \cite{Padmanabhan:2010qr,Li:2010cj}.
As is well known, the Friedman equation can be deduced from semi-Newtonian physics. Thus it ensues from the
above arguments as shown by reference \cite{Cai:2010hk}. Reference \cite{Verlinde:2010hp} has also inspired modifications to the
cosmic expansion laws. 
The purpose of the present paper is to
uncover implications of such modifications. Two approaches are investigated\footnote{Other approaches include considering two holographic screens \cite{Cai:2010zw} or Debye modifications of the temperature dependence \cite{Gao:2010fw}. Cosmological implications of the doublescreen model \cite{Cai:2010kp} and of the Debye model \cite{Wei:2010am, Wei:2010wwa} have also been considered.}. One set of corrections to the Friedman equations is motivated by the possible connection of the surface terms in the gravitational action to the holographic entropy.
Reference \cite{Danielsson:2010uy} noted that (at the present level of the formulation) this is equivalent to introducing
sources to the continuity equations, previously considered in the trans-Planckian context. Thus
non-conservation of energy is implied. In another approach, the derivation of the Friedman
equation as an entropic force from basic thermodynamic principles is generalized by taking into account loop
corrections to the entropy-area law \cite{Sheykhi:2010wm}. The result is corroborated by its accordance with
previous considerations \cite{Cai:2008gw}, and is consistent with energy conservation though yet lacks a
covariant formulation. While it might seem preliminary to investigate in detail the predictions of these models
whose foundations, at the present stage, are rather heuristic, we believe it is useful to explore the generic
consequences such models may have. Knowing about the possible form of viable extensions to our standard
Friedmannian picture along the lines of reference \cite{Dvali:2003rk} can shed light on the way towards more rigorous
derivation of the
effective entropic cosmology, and on the prospects of eventually testing the above cited ideas by cosmological observations.

An encouraging result in this respect is that the viable cosmologies in a realisations of both the quite
different approaches
we focus on, possess an identical expansion rate (in the simplest but relevant setting of a universe filled by a single fluid dominated cosmology). It is also interesting that at high curvatures this expansion rate reduces to a constant. Thus not a big bang singularity, but instead inflation is found in the past. We present a simple argument why this inflation avoids the no-go theorem formulated in reference \cite{Li:2010bc}, which indeed is valid for material sources violating the strong energy condition. We also find that the higher curvature corrections, motivated by quantum corrections, are not viable as they lack a consistent low-energy limit. Furthermore, we impose  bounds on the unknown parameters of the models by considering the scale of inflation,  big bang nucleosynthesis (BBN) and from the effects of a modified behaviour of dark matter in the post-recombination universe.

Only phenomenologically motivated, but an interesting case is a monomial correction to the
area-entropy law. Such can result in acceleration without dark energy with a good fit to the data. We perform a
full Markov Chain Monte Carlo (MCMC) likelihood analysis exploiting astronomical data from baryon acoustic oscillations \cite{bao7},
supernovae \cite{union2} and cosmic microwave background \cite{wmap}. A slightly closed universe turns then out to be
preferred by the data, unlike in the standard $\Lambda$CDM model. We consider also the evolution perturbations,
which is determined uniquely if the Jebsen-Birkhoff law is valid. A characteristic feature is then the growth of
gravitational potentials in conjunction with the modified growth of overdensities. We also show that the visible
universe bounded by the last scattering surface is less entropic than a black hole the enclosed volume would form. This
consistency check proposed in reference \cite{Frampton:2010xu} can also be employed to constrain entropic cosmologies.

The surface term approach is discussed in section \ref{surface}. In section \ref{loop} we consider the
implications of a specific form of area-entropy law motivated by quantum gravity, and in section
\ref{generalized} we explore a phenomenological power-law parametrization of this law. The perturbation
evolution and constraints from more theoretical considerations are discussed in section \ref{comments}.
Each of these sections can be read independently of the others.
Finally, the results we obtained are concisely summarized in section \ref{conclusions}.

\section{Modifications from surface terms}
\label{surface}

Easson, Frampton and Smoot recently argued that extra terms should be added to the acceleration equation for the
scale factor. 
This was discussed from various points of view, in particular
it was conjectured that the additional terms can stem from the usually neglected surface terms in the
gravitational action. Present acceleration of the universe \cite{Easson:2010xf} and inflation
\cite{Easson:2010av} were proposed to be explained by the presence of these terms without introducing new fields. This is obviously an exciting prospect warranting closer inspection.

Slightly different versions of the acceleration equation were introduced in both of the above mentioned two papers. The following parametrization of the two Friedman  equations encompass all those versions and their combinations\footnote{We will not adress the case in which the two equations degenerate to one by a particular choice of the parameters.}:
\ba \label{efs_friedman}
H^2 & = & \frac{8\pi G}{3}\rho + \al_1H^2+\al_2\dot{H}+ 8\pi G\al_3 H^4\,, \label{efs_1} \\
\dot{H} + H^2 & = & -\frac{4\pi G}{3}(1+3w)\rho + \bt_1 H^2 + \bt_2 \dot{H} + 8\pi G\bt_3 H^4\,. \nonumber
\ea
The six coefficients $\al_i$, $\bt_i$ are dimensionless for all $i=1,2,3$.
The extrinsic curvature at the surface was argued to result in something like $\al_1=\bt_1=3/2\pi$ and $\al_2=\bt_2=3/4\pi$ and quantum corrections in nonzero $\bt_3$ \cite{Easson:2010av,Easson:2010xf}.
The equations imply that
\be \label{efs_3}
\frac{d H}{dN} = H\lp 8\pi G c_2 H^2 -c_1\rp \,,
\ee
where $N=\log(a)$ is the e-folding time and
\ba \label{c_i}
c_1 &\equiv& \frac{(3-\al_1)(1+w)-2\bt_1}{2-\al_2(1+3w)-2\bt_2}\,, \\ c_2 &=& \frac{\al_3(1+3w)+2\bt_3}{2-\al_2(1+3w)-2\bt_2}\,.
\ea
Note that $c_1$ is proportional to the lower order corrections, and $c_2$ is proportional to the higher order contributions $\sim H^4$. The information lost by having only one differential equation in (\ref{efs_3}) should be compensated by imposing boundary conditions from (\ref{efs_friedman}) to its solutions.
In the general case of multiple fluids, the model does not uniquely determine how to the EoS (equation of
state) $w$ evolves. The
reason is that the two equations (\ref{efs_1}) result in only one (non)conservation equation for the total density, and we have no unique prescription how the relative densities behave if the total density consists of a mixture of fluids. From the viewpoint of reference \cite{Danielsson:2010uy}, the source terms for the individual fluids are undetermined. However, the most relevant special case of a single-fluid dominated universe allows an exact solution where these ambiguities are absent.

\subsection{Single fluid}

In the case that $w$ is a constant, Eq.(\ref{efs_3}) can be easily solved:
\ba \label{efs_4}
H^2(a) &= &\frac{c_1}{8\pi G\lb c_0\lp\frac{a}{a_0}\rp^{2c_1} + c_2\rb}\,, \\ c_0 &=& \frac{1}{32\pi^2 G^2(1+w)\rho_0}\,.
\ea
We chose the integration constant $c_0$ in such a way that when the entropic corrections vanish, we recover the standard Hubble law.
It is now clear that though there are six unknown factors in (\ref{efs_1}) we cannot derive from first principles, the cosmological implications are rather unambiguous (some degeneracies are broken for evolving $w$ as we also explicitly see below), and can be encoded in the two numbers $c_1$ and $c_2$. Thus it is both feasible and meaningful to constrain them, despite our considerable ignorance of the more fundamental starting point. From the form (\ref{efs_4}) it is also transparent that as $a\rightarrow 0$, we have de Sitter solution: so the model indeed predicts inflation. As the scale factor grows, (nearly) standard evolution is recovered: so it is also simple to see the present versions of the model don't provide dark energy. At early times, $w=1/3$, we can obtain constraints from BBN, and from the inflationary scale by estimating the amplitude of fluctuations. Both the scaling modification $c_1$ and the constant term $c_2$ can be bounded. At late times, $w=0$, we can obtain constraints at least from the modified scaling law for dust, $c_1$. Before this let us however consider obtaining the present acceleration.

\subsection{Adding a cosmological constant}

As we need acceleration at late times, we need a $\Lambda$-term to accelerate the universe. Then (\ref{efs_3}) generalizes to
\ba \label{efs_l}
\frac{d H}{dN} &=& H\lp 8\pi G c_2 H^2 -c_1 \rp + c_3\frac{\Lambda}{H} \,, \\ c_3 &=& \frac{1+w}{2-\al_2(1+3w)+2\bt_2}\,.
\ea
This equation is solved by
\ba \label{efs_h}
H^2=\frac{c_1}{4\pi G c_2} +  \frac{\sqrt{32\pi G c_2c_3 \La - c_1^2}}{16\pi G c_2}\tanh\lp\sqrt{32\pi G c_2c_3 \La - c_1^2}(N-N_0)\rp 
\ea
Hence, the form of the Friedman equation is quite completely different from the usual. This is due to the
nonlinearity of stemming from the presence the higher curvature corrections. At low curvatures, their effect
doesn't disappear as the most naive expectation would be. In fact, the limit $c_2\rightarrow 0$ is not defined
for Eq.(\ref{efs_h}). At asymptotically late times, this reduces to the de Sitter solution, but the preceding
evolution may not approximate standard cosmology. To cure this, we suggest introducing a suitable source term for
the $\Lambda$-term, which then becomes dynamical rather than a cosmological constant. This can also improve the phenomenological viability of the model, since the constraints from modified matter scaling (see below) would not hold.

\subsection{Prescription I: constant $\Lambda$}

Thus we saw that the higher curvature corrections together are not compatible with a cosmological constant in a viable cosmology.
Let us therefore consider the case that the higher curvature corrections given by $c_2$ vanish, but allow a $\Lambda$ term to accelerate the universe. Then the behavior of the Hubble rate is just what is expected from Eq.(\ref{efs_4}). Now (\ref{efs_l}) is solved by
\be
H^2(a) = \frac{8\pi G}{3}\rho_0 a^{-2c_1} + \frac{1}{3-\frac{2}{1+w}\bt_1}\La\,.
\ee
The integration constant $\rho_0$ corresponds to the renormalised energy density at $a=1$. Similarly, the cosmological constant is slightly ''dressed''. The conclusion is that the observable effect to the expansion is the modified scaling of matter density. Below we consider the cosmological bounds on such scaling.

\subsubsection{Early universe constraints}
\label{efs_cons}

First we consider the constraints from early universe. During radiation domination, Eq. (\ref{efs_4}) becomes
\be
H^2=8\pi G \rho \frac{\frac{1}{2}c_1}{a^{2(c_1-2)} + 32\pi G^2 c_2\rho}\,.
\ee
From this we see that the variation effective Newton's constant is
\be
\delta G_{eff}/G \approx (\frac{1}{2}c_1-1)(1-4c_1\log{a})  - 16\pi^2 G^2 c_1 c_2 \rho\,.
\ee
This variation can be bounded by requiring successful BBN. For instance, reference \cite{Bambi:2005fi} derived that
$\delta G_{eff} = 0.09^{+0.22}_{-0.19}$. The radiation energy density is given by
\be \label{radiation}
\rho=g_*\frac{\pi^2}{30}T^4\,,
\ee
where we use for the number of effective relativistic degrees of freedom $g_*$ at the nucleosynthesis temperature $T \sim 1$ MeV the value $g_* = 10.75$. Plugging in the numbers, we obtain
\be
-3.5\cdot 10^{-3} < 2-c_1 < 1.1\cdot 10^{-3} \,.
\ee
\be
-2 \cdot 10^{84} < c_2 < 6\cdot 10^{84} \,.
\ee
Because of the tremendous hierarchy between the Planck and the BBN scale there is a very poor constraint on the high curvature corrections $c_2$. This can be also written as $2\sqrt{\pi G} c_2^{1/4} < 457.6 \, GeV^{-1}$ by restoring the dimensions.

If inflation is considered to be driven by the entropic corrections, we can estimate their magnitude from the amplitude of perturbations observed in CMB. Amplitude of the spectrum of quantum fluctuations of massless fields is expected to be given by the ratio
\be
 \langle\delta\phi\delta\phi\rangle=\frac{8\pi G H^2}{\epsilon} \sim 10^{-10}\,,
\ee
 where $\epsilon$ is the slow-roll parameter and the right hand side is determined from observations. The
spectral index as determined from observations gives $\epsilon \sim \mathcal{O}(0.01)$. Since at early times
equation
(\ref{efs_4}) predicts (nearly) exponential expansion with the Hubble rate $8\pi G H^2 = c_1/c_2$, and we know
that $c_1$ must be of order one, successful generation of observed fluctuations from entropic inflation suggest
that $c_2 \sim 10^{12}$. To obtain inflation at the GUT scale for example, we have to consider a very large amplification of the Planck-scale suppressed effect of $c_2 \sim M_P/M_{GUT}$.
This estimate is much more tentative than the other we impose, since it depends on the physics of inflation, that are not well established even in standard cosmology. Reference \cite{Wang:2010jm} discussed holographic view of inflation and the interpretation of quantum fluctuations as thermal fluctuations on the screen.

\subsubsection{Late universe constraints}

The modified scaling law of dark matter can be used to impose tight bounds from the late universe. This has been explored in reference
\cite{Muller:2004yb}, who derived constraints on the EoS for dark matter, taking into
account experimental data both on the background and on the perturbations. Adopting the prescription where the
Newton frame sound speed vanishes\footnote{Ref.\cite {Muller:2004yb} uses slightly nonstandard nomenclature for
the
sound speeds. Usually $\dot{p}/\dot{\rho}$ is referred to as the adiabatic sound speed. The gauge-dependent
quantity $\delta p/\delta\rho$, when evaluated in the rest frame of the fluid, gives the sound speed according
to the usual definition, see e.g. \cite{Bean:2003fb, Koivisto:2005mm, Mota:2007sz}.} we can
translate the result into our case as:
\be
-8.78 \cdot 10^{-3} < 2c_1-3 < 1.86 \cdot 10^{-3}\,.
\ee
for 99.7\% C.L. bounds. Reference \cite {Muller:2004yb} took into account the full CMB and LSS data. However, as we cannot deduce the perturbation
evolution in these models unambiguosly, it is useful to consider constraints ensuing solely from
background expansion. It turns out that by including the latest data on SNeIa, BAO and CMB, the reached precision is only slightly lower. The result is shown in Fig. \ref{c1} and corresponds to the bounds
\be
-17.28 \cdot 10^{-3} < 2c_1-3 < 20.50 \cdot 10^{-3} \,. 
\ee
at 99.7\% C.L. As proposed in reference \cite{Easson:2010av}, a more complete version of the model could also be constrained by the precision data on the equivalence principle.

\FIGURE{
\includegraphics[width=.6 \textwidth]{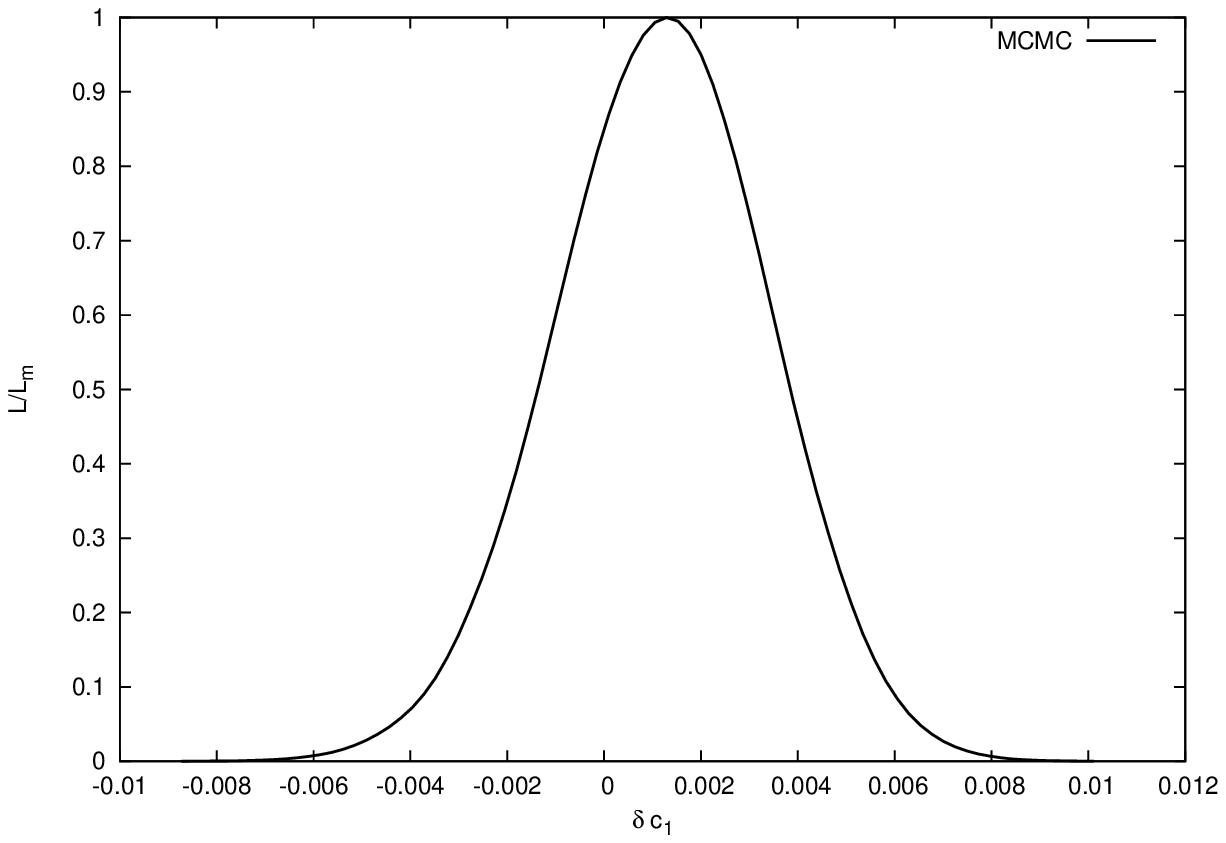}
\caption{\label{c1}Constraints on the modified equation of state for CDM and baryons scaling law by using SNe,
BAO and
CMB distance priors. Radiation has been assumed to follow the usual scaling. Only very slight deviations from the usual scaling law are allowed.}
}

\subsection{Prescription II: dynamical $\Lambda$}
\label{pres2}

As mentioned above, one can also consider the case that matter continuity equation is not violated. Then the $\Lambda$-term must be responsible for the non-conservation in a consistent system. The Hubble law can be derived analogously to the above cases and one readily finds that it now has the form
\be
H^2=\frac{8\pi G \rho}{3\lp 1 + \frac{2(\al_1-\bt_1)}{3(1+w)}\rp} + \Lambda_0 a^{2(\al_1-\bt_1)}\,.
\ee
So the $\Lambda$-term acquires a dynamical behavior. In case $\al_1-\bt_1<0$ this would help with the cosmological constant problems, since one could consider initial large values for $\Lambda$, which has diluted to the presently observed scale. Note that this is different from usual dark energy approach, where $\Lambda$ is tuned to zero (or in any case to an even smaller than the value consistent with observations) and then a new dynamical component is added to explain the acceleration. Reference \cite{Casadio:2010fs} have also derived this result, which can be equivalently arrived at by imposing only the second Friedman equation in (\ref{efs_friedman}). The first one then follows by integration, and the dynamical $\La$ can be viewed as an integration constant.

\subsubsection{Constraints}

Now the BBN constraint for the effective gravitational constant gives
\be
-0.62 < \bt_1-\al_1  < 0.20 \,. 
\ee
From WMAP7 measurements on the equation of state of dark energy, combined with other cosmological data
\cite{Komatsu:2010fb}, we obtain an even tighter bound,
\be
-0.05 < \bt_1-\al_1 < 0.11 \,. 
\ee
This is in qualitative agreement with reference \cite{Casadio:2010fs}, where the entropic corrections were bounded with the CMB acoustic scale.

\section{Modifications from quantum corrections to the entropy-area law }
\label{loop}

There is evidence from string theory and from loop quantum gravity that the two leading quantum corrections to the area entropy-law are proportional to the logarithm and the inverse of the area \cite{Peet:2000hn,Carlip:2008rk}.

 Reference \cite{Sheykhi:2010wm} derived the Friedman equation from an underlying entropic force taking into account quantum corrections to the entropy formula. We slightly generalise his end result Eq.(25) by allowing multiple fluids (labeled $i$) with constant equations of state $w_i$ and a cosmological constant:
\ba
H^2 &+& \frac{k}{a^2}  =  \sum_{i}\lb\frac{8\pi G}{3} - \frac{8\bt(1+3w_i) G^2}{9(1+w_i)}\lp H^2+\frac{k}{a^2}\rp
\nonumber\right. \\ &-& \left. \frac{2{\gamma} (1+3w_i) G^3}{3\pi(5+3w_i)}\lp H^2+\frac{k}{a^2}\rp^2 \rb \rho_i  \\
& + & \frac{\Lambda}{3}\lb 1-\frac{\bt}{\pi}G(H^2+\frac{k}{a^2})\log{a} + \frac{{\gamma}}{4\pi^2}G^2(H^2+\frac{k}{a^2})^2\rb.\nonumber
\ea
Again there is a problem recovering usual evolution at low curvature if we include the high curvature correction proportional to ${\gamma}$.
This can be seen easily. Defining the shorthand notations
\ba
S_\beta & \equiv & -\frac{8\bt G^2}{9}\sum_{i}\frac{1+3w_i}{1+w_i}\rho_i - \bt\frac{G\Lambda }{3\pi}\log{a} \,, \\
S_{\gamma}      & \equiv & -\frac{2{\gamma} G^2}{3\pi}\sum_{i}\frac{1+3w_i}{5+3w_i}\rho_i + {\gamma}\frac{G\Lambda}{12\pi^2}\,,
\ea
the solution(s)  for the Hubble rate may be written as
\be
H^2+\frac{k}{a^2} = \frac{1-S_\bt}{2G S_{\gamma}} \pm
\frac{\sqrt{1-12\pi G^2\rho S_{\gamma}}}{2G S_{\gamma}}\,,
\ee
where the total matter density is denoted by $\rho=\sum \rho_i$.
It is obvious that at the limit where the corrections tend to zero, we do not recover standard cosmological evolution. Thus the higher order corrections here suffer from a similar graceful exit problem as we encountered in the previous section.

Therefore we set ${\gamma}=0$ and consider only effect of the leading logarithm correction to the entropy, proportional to $\bt$ . The solution for the Hubble rate can then be written as, neglecting the cosmological constant,
\be \label{s_1}
H^2 + \frac{k}{a^2} = \frac{8\pi G}{3}\rho\lb 1 + \frac{8\bt G^2}{9} \sum_i \lp \frac{1+3w_i}{1+w_i}\rp \rho_i \rb^{-1}\,.
\ee
Thus, the corrections occur near the Planck scale. If $\bt$ is large enough this can support inflation since the RHS tends to a constant when matter is relativistic and $w_i=1/3$ for all species $i$. Again, we can constrain this from the effective $G$ at BBN. It is interesting to note that the form of the entropic Friedman equation assumes the same form as in the previous section, where the derivation was quite different, the underlying physical assumptions leading to a (non-)conservation and apparently different form of the force law. The only difference between (\ref{efs_4}) and (\ref{s_1}) is that in the case where the matter conservation laws are modified, (\ref{efs_1}) can describe also a slightly modified scaling of the matter density. This may be regarded as support for the robustness of the prediction of the expansion law in entropic cosmologies.

\subsection{Constraints}

 From Eq.(\ref{s_1}), the effective variation of the Newton's constant is now given by
 \be
 \delta G_N/G = \frac{1}{1+\frac{4\bt G^2}{3}\rho_{BBN}}-1 \approx - \frac{4}{3}\bt G^2 \rho\,.
 \ee
 Using Eq.(\ref{radiation}) for the radiation density at nucleosynthesis and proceeding analogously to section \ref{efs_cons}, we find that the BBN bound on the magnitude constant $\bt$ is $|\bt| \lesssim 2.7 \cdot 10^{35}$, translating to
 \be
 |\sqrt{8\pi G \bt}| < 0.0042 \quad 1/GeV\,.
 \ee
Again it is clear that BBN is not efficient to constrain the corrections. Furthermore, since the scale of inflation is below the Planck scale, we have to consider very large values of $\bt$. From considerations of loop quantum gravity and string theory however, the natural value for $\bt$ is of order one. Considering such values, inflation takes place at the Planck scale - where we cannot trust the perturbatively entropy-area law, which can be expected to hold only at the limit of large horizon size. In fact, e.g. \cite{Nicolini:2010nb} has argued that the description of spacetime as a differential manifold may be justified only asymptotically at macroscopic length scales.

\section{Dark energy from generalized entropy-area law?}
\label{generalized}

In the following we consider the possibility of infrared modifications to the large-scale behavior of gravity. Such can ensue from corrections to the $S\sim A$ relation that grow faster than $A$. Among such is the volume correction that scales as $\sim A^{3/2}$. Interestingly, corrections of this type imply, within the entropic interpretation of gravity, a modified Newton's law which may explain the galactic rotation curves without resorting to dark matter. This has been shown by reference \cite{Modesto:2010rm}. We can then, effectively, generate MOND \cite{Milgrom:1983ca,Bourliot:2006ig} at the galactic scales. This motivates us to study whether we may generate modified gravity at the largest scales in such a way that we would avoid the introduction of dark energy field or a cosmological constant.

For this purpose, we consider the area-entropy law of the form
\be
S=\frac{A}{4\ell_P^2} + s(A)\,,
\ee
where the function $s(A)$ represents the quantum corrections and
$\ell_P^2=G\hbar/c^3$. We assume that $A=QN$, where $Q$ is a constant to
be determined, and that the entropy changes by one fundamental unit
(corresponding to unit change in the number of bits on the screen with
radius $R$) when $\Delta r = \eta \hbar/(mc)$, $r$ being the comoving
radial coordinate. Then the first law of thermodynamics together with
the equipartition of energy leads to the modified Newtonian law of
gravitation\footnote{In Ref.\cite{Zhang:2010hi} it was instead assumed
that the number of bits is directly proportional to entropy, which is
not compatible with our assumption $A=QN$. From the former assumption
follows instead $F=-G M m/(R^2 + \ell_P^2s(A)/\pi)$.}
\be
F=-\frac{Q^2c^3 M m}{2\pi k_B \hbar\eta R^2}\left(\frac{1}{\ell_P^2}+\frac{\partial s}{\partial A} \right) = -\frac{G M m}{R^2}\lp 1 + 4\ell_P^2\frac{\partial s}{\partial A}\rp\,,
\ee
where in the second equality we made the identification $Q^2=8\pi k_B \eta \ell_P^4$. Let us assume the power-law correction
\be \label{powerlaw}
s(A) = \frac{4\pi\sa}{n}\lp\frac{a r}{\ell_P}\rp^{2n} \sim A^n\,.
\ee
This type of parametrization for entropic gravity effects has been recently considered by other authors \cite{Sheykhi:2010yq}.
Taking into account that in the cosmological context the active gravitational mass is given by the Tolman-Komar mass
(\ref{tk_frw}), and that the $R=ar=1/\sqrt{H^2+k/a^2}$, we obtain the Friedman equation
\ba \label{friedmann}
H^2  +  \frac{k}{a^2} =  \frac{8\pi G}{3}\sum_i \Bigg[ 1  + 
\sa\frac{(1+3w_i)}{1+3w_i-2n}\lp\frac{1}{\ell_P^2(H^2+\frac{k}{a^2})}\rp^{n-1}\Bigg]\rho_i \,.
\ea
Not surprisingly, the possible infrared corrections, $n>1$, are precisely those which could be significant in cosmology at late times. The nonperturbative form of $s(A)$ is of course is unknown, but the volume correction is known to be given by $n=3/2$, so $n>1$ is not something to exclude a priori. In the following we explore how cosmological observations constrain the parametrization (\ref{powerlaw}).

In the flat case, if the energy density is dominated by a fluid with the EoS $w$ and the corrections
dominate over the standard term in (\ref{friedmann}), the expansion is described by the effective EoS
\be \label{weff}
w_{eff} = \frac{1+w}{n}-1\,.
\ee
Thus a matter dominated universe accelerates given $n>3$. With larger $n$, the effective EoS is
more negative, but phantom expansion can be achieved only when $w$ is itself negative. The exact evolution of $w_{eff}$ including the effects of possible spatial curvature is shown in Figure \ref{weff_fig}.

\FIGURE{
\includegraphics[width=.6 \textwidth]{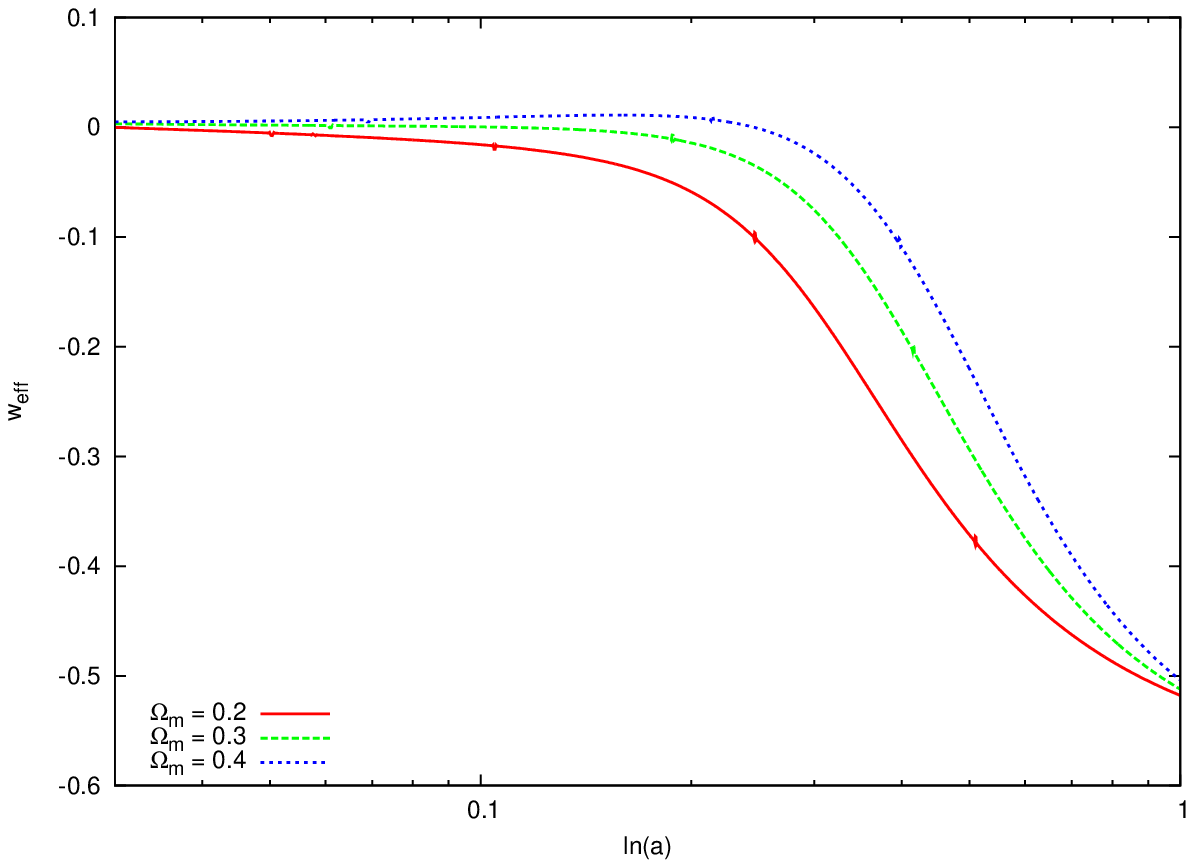}
\caption{Effective equation of state for the generalized entropy-area law. The lines correspond to an open, flat and closed universe with $\Omega_S=0.7$ and $n=2.5$. Equation (\ref{weff}) would give $w=-0.6$ for this case.}\label{weff_fig}
}

\subsection{Observational constraints} \label{observational-section}

In order to obtain the bounds on the parameters arising from the
modified Friedman equation, a suitably modified version of CMBeasy
\cite{cmbeasy} was employed together with a MCMC code, taking into
account astronomical data from baryon acoustic oscillations \cite{bao7}, supernovae
\cite{union2} and cosmic microwave background \cite{wmap}. The results are displayed in Figure \ref{mcmc1} and Table \ref{results}.
Due to the geometric nature of the modifications, a possible curvature of the
spatial sections was allowed. This revealed a preference towards slightly closed universes, which might be due to the appearance of $k$ in the r.h.s. of (\ref{friedmann}).
Relatively lower values of $n$ are favoured with respect to higher ones because higher values reproduce a total equation of state which is too close to $-1$. Note also the existence of degeneracies in the individual datasets, which are broken by the combined constraints.

\FIGURE{
\includegraphics[width=.49 \textwidth]{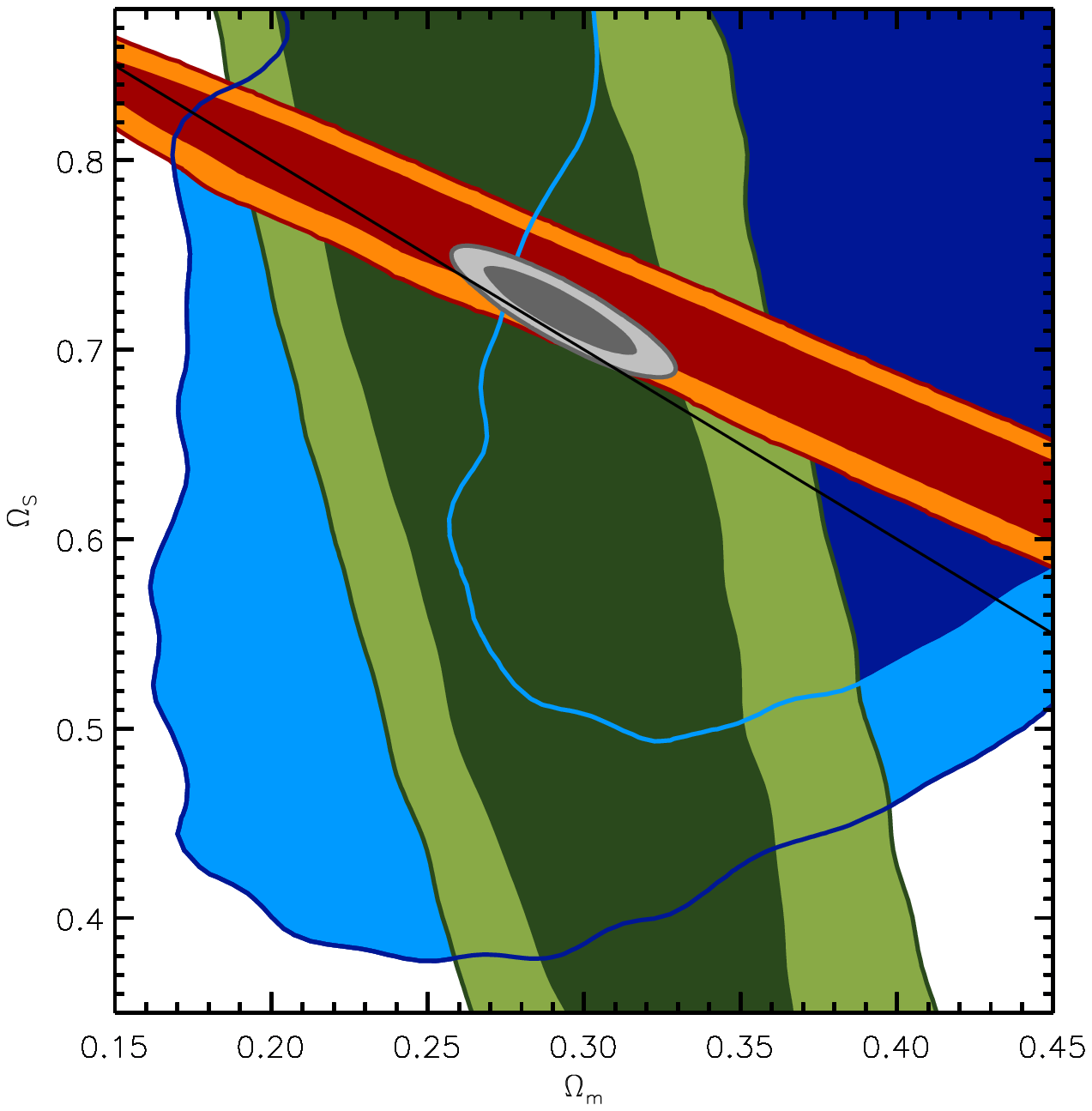}
\includegraphics[width=.49 \textwidth]{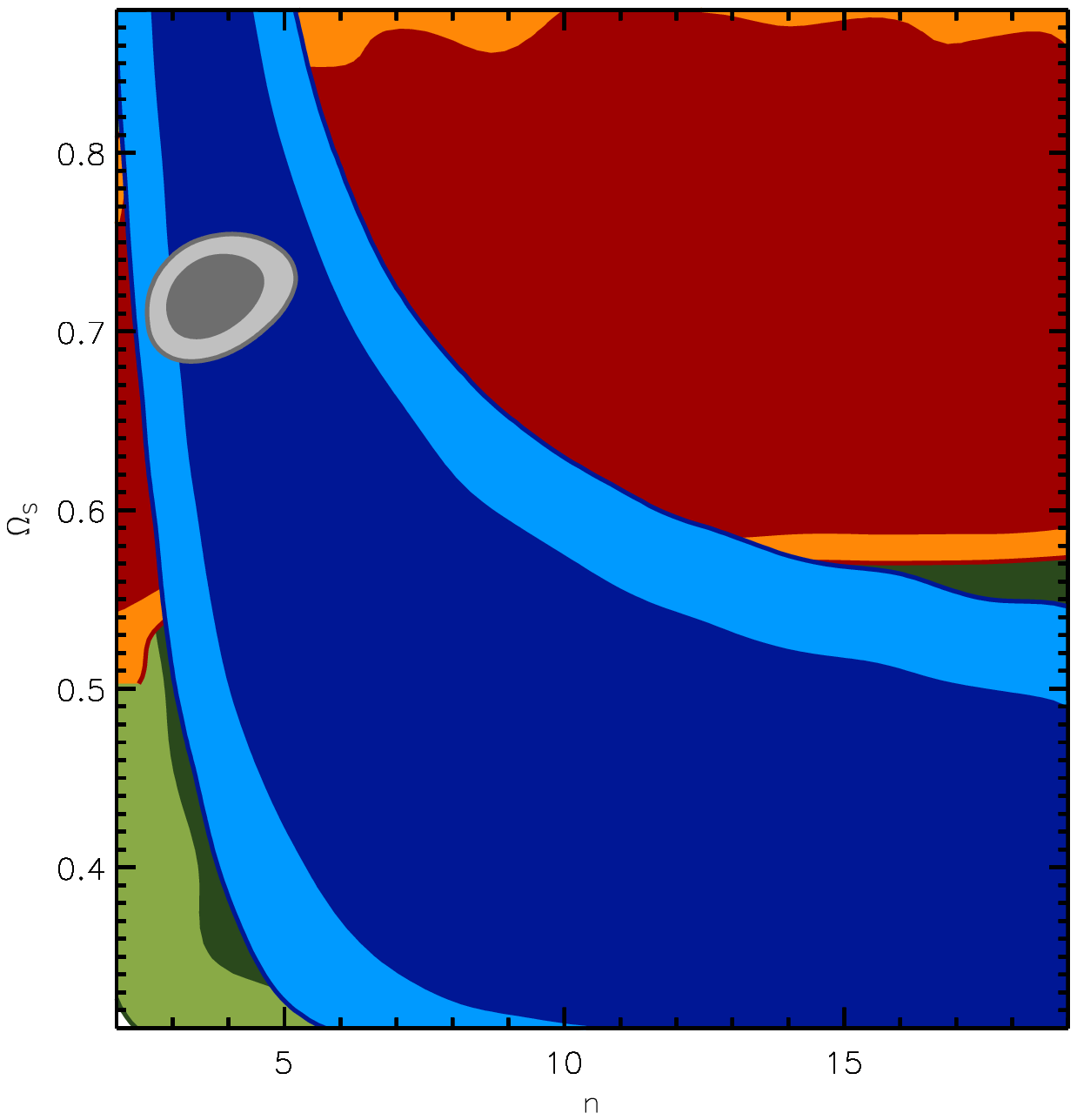}
\caption{Bounds on parameters for the generalized area-entropy law using Sne (blue), CMB (orange), BAO (green)
and the three combined datasets (gray). Note that closed universes ($\Omega_m + \Omega_S > 1$) are slightly preferred in this model. Entropic corrections gave $\chi^2_S = 533.28$, slightly higher than the value obtained for a similar MCMC for $\Lambda$CDM with $\chi^2_\Lambda = 532.34$.}\label{mcmc1}
}

\TABLE{
\begin{tabular}{c  c c c c   }
 &  All & BAO & CMB & SNe    \\
\hline \\[-1pt]
$h$ & $ 0.68 \pm {0.02}$ & $ - $ & $ 0.58 \pm 0.06 $ & $  - $   \\[4pt]
$n$ & $ 3.8 \pm 0.7 $ & $ >2.9 $ & $ - $ & $  > 3.0 $  \\[3pt]
$\Omega_S $ & $ 0.69\pm 0.02 $  & $-$ & $ 0.71 \pm 0.12 $ & $ 0.76 \pm 0.21 $     \\[3pt]
$\Omega_m $ & $ 0.29\pm 0.02 $ & $ 0.27 \pm 0.05 $ & $ 0.33 \pm 0.13 $ & $ 0.39 \pm 0.10 $    \\[3pt]
\hline \\[-1pt]
$\Omega_k $ & $ 0.012 \pm 0.005 $  & $ - $ & $ -0.04 \pm 0.05  $ & $ -0.16 \pm 0.24 $    \\[3pt]
\hline
\end{tabular}
\caption{Maximum likelihood values and 1 sigma error bars from the constraints of Section \ref{observational-section}. \label{results}}
}

Table \ref{results2} shows the results of model comparison with $\Lambda$CDM including the Bayesian and the Akaike criteria given by $-2\log L_{max}+ p \log d$ and $-2\log L_{max}+ 2 p$ respectively, with $p$ the number of free parameters and $d$ the number of experimental data points. Eventhough a cosmological constant is favoured in all cases, the values of $\chi^2$ are very similar and most of the difference in these cases is due to the additional parameter $n$ in the entropy-corrected model.

\TABLE{
\hspace{2.5cm} 
\begin{tabular}{c  c c  }
&  $S(A)$ & $\Lambda$ \\[4pt] \hline
$\chi^2$ & 533.28 & 532.34 \\[4pt]
$\chi^2/d.o.f$ & 0.956 & 0.952 \\[4pt]
Bayesian & 558.61 & 551.31 \\[4pt]
Akaike & 541.28 & 538.32 \\ \hline
\end{tabular}
\hspace{2.5cm} 
\caption{Model comparison according to different criteria. \label{results2}}
}


\section{General comments on entropic cosmologies}
\label{comments}

\subsection{On the evolution of perturbations}
\label{perts}

Reference \cite{Lue:2003ky} have shown that by assuming the Jebsen-Birkhoff theorem \cite{VojeJohansen:2005nd} it is
possible to deduce the evolution of the spherical overdensities in a dust-filled universe given the
background evolution. This is equivalent to a brane-motivated set-up where an effective energy density (in our
case with fractional density $\Omega_s$) evolves adiabatically with cold dark matter as shown in Ref. \cite{Koivisto:2004ne}.
It is not clear to us whether the entropic gravity obeys the Jebsen-Birkhoff theorem. Though this seems in line
with the equipartition principle, in the presence of corrections to the $S\sim A$ law the case is more
nontrivial. Therefore we didn't include the constraints from perturbations into the likelihood analysis above.
However, in the spirit of reference \cite{Lue:2003ky} (where the approach was developed to study DGP and related models) we
take this as a first approximation to gain insight into the clustering of matter in entropic cosmology. The
perturbation evolution equation can be derived by tracking the surface of a star in
a Schwarzchild metric embedded in the background of FRW where the expansion is given by some gravity
theory deviating from Einstein's GR.

Consider the curved background
\be \label{cmetric}
ds^2=-dt^2+a^2(t)\lp \frac{d\lambda^2}{1-K\lambda^2}+\lambda^2 d\Omega^2 \rp\,.
\ee
We want to match this with the Schwarzchild-like metric
\be \label{smetric}
ds^2 = -g_{00}(r)dT^2+g_{rr}(r)dr^2+r^2 d\Omega^2\,.
\ee
For this purpose consider the coordinate transformation $r=r(t,\lambda)$, $T=T(t,\lambda)$ which allows to rewrite (\ref{smetric}) in the form
\be
ds^2 = -N^2(t,\lambda)dt^2+a^2(t)\frac{d\lambda^2}{1-K\lambda^2}+r^2 d\Omega^2\,.
\ee
This implies the following conditions are satisfied:
\ba
-g_{00}\dot{T}^2+g_{rr}\dot{r}^2& = & -N^2\,, \label{cond1} \\
-g_{00}\dot{T}T'+g_{rr}\dot{r}r'& = &  0 \,, \label{cond2} \\
-g_{00}(T')^2+g_{rr}(r')^2 & = & \frac{a^2}{1-K\lambda^2}\,. \label{cond3}
\ea
Here prime indicates derivative wrt $\lambda$, and an overdot derivative wrt $t$.
We consider a spherical object whose interior expands like (\ref{cmetric}), and match the
boundary at $\lambda=\lambda_*$ smoothly with the exterior metric (\ref{smetric}).
This requires
\ba
r(t,\lambda_*)& = & a(t)\lambda_*\,, \label{match1} \\
r'(t,\lambda_*)& = & a(t)\,, \label{match2} \\
N(t,\lambda_*) & = & 1\,, \label{match3} \\
N'(t,\lambda_*)& = & 0\,. \label{match4}
\ea
With some algebra employing equations (\ref{cond1}-\ref{match3}) we infer that
\be \label{grr}
g_{rr}(r(t,\lambda_*))=\frac{1}{1-\lambda_*^2\lp \dot{a}^2 + K\rp}\,.
\ee
One obtains directly from (\ref{match1}) that the radius at the boundary $\lambda_*$ satisfies
\be \label{evol_r}
\ddot{r} = r(H^2 + \dot{H})\,.
\ee
Since the top-hat overdensity $\delta(t)$ contained within radius $r$ in the background density $\rho_M$ is $(1+\delta) \sim 1/(\rho_M r^3)$, we can recast (\ref{evol_r}) into an evolution equation for $\delta$. At linear order, we obtain
\be \label{lue}
\ddot{\delta} + 2H\dot{\delta} = \left(2\dot{H}+\frac{\ddot{H}}{H}\right)\delta\,.
\ee
In contrast, when the same background expansion is due to a smooth dark energy component, the growth of
perturbations is governed by
\be \label{smooth}
\ddot{\delta} + 2H\dot{\delta} = 4\pi G \rho_M\delta\,.
\ee
The difference is thus only the source term in the RHS of Eq.(\ref{lue}) due to the clumpiness of the effective
fluid, whereas with smooth dark energy only the matter density acts as a gravitational source in the RHS of
(\ref{smooth}).
This also determines the behaviour of the metric perturbations, which can be probed by various observations, in particular
weak lensing and ISW and its correlations. In the longitudinal gauge the scalar perturbations are parameterized by
\be \label{met1}
ds^2=-(1+2\Psi)dt^2+a^2(t)(1+2\Phi)\lb \frac{d\lambda^2}{1-K\lambda^2} + d\lambda^2d\Omega \rb \,.
\ee
Now we know that in an overdense region the line element can be written as
\be \label{met2}
ds^2=-d\tilde{t}^2+\tilde{a}^2(\tilde{t})\lb \frac{d\tilde{\lambda}^2}{1-\tilde{K}\tilde{\lambda}^2} + d\tilde{\lambda}^2d\Omega \rb \,,
\ee
where the curvature $\tilde{K}$ is associated with the overdensity and $\tilde{a}^3=(1+\delta)a^3$. Analogously to Eq.(\ref{grr}), one may now infer that
\be
g^{-1}_{rr} = 1-\dot{r}^2-\tilde{\lambda}^2_* \tilde{K}\,.
\ee
It follows that
\be
\tilde{K}-K=-\frac{2}{3}\lp \dot{H}\delta + H\dot{\delta} \rp\,.
\ee
What remains to do is a coordinate transformation that brings (\ref{met2}) into the form (\ref{met1}). With the top-hat profile for the overdensity, we can then identify the coefficients $\Phi$ and $\Psi$. Follwing reference \cite{Lue:2003ky} one may then find that
\ba
\frac{\nabla^2}{a^2}\Phi & = & \dot{H}\delta\,, \label{lue2} \\
\frac{\nabla^2}{a^2}\Psi & = & -\lp 2\dot{H} + \frac{\ddot{H}}{H}\rp\delta \label{lue3} \,.
\ea
This shows that the entropic corrections forces the gravitational potential unequal and thereby producing effective anisotropic stress. This is an interesting prediction as it allows to distinguish the possible entropic origin of acceleration from for instance scalar field models.

\FIGURE{
\includegraphics[width=.49 \textwidth]{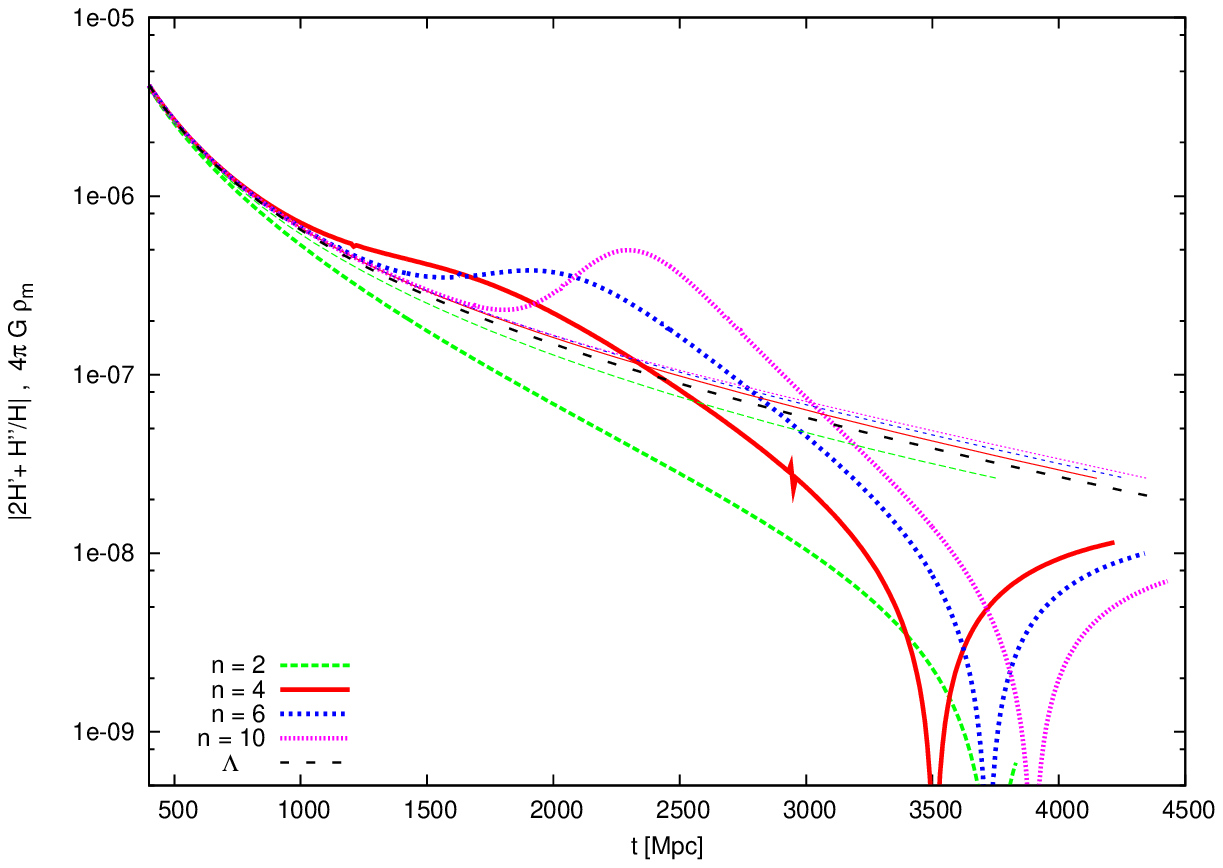}
\includegraphics[width=.49 \textwidth]{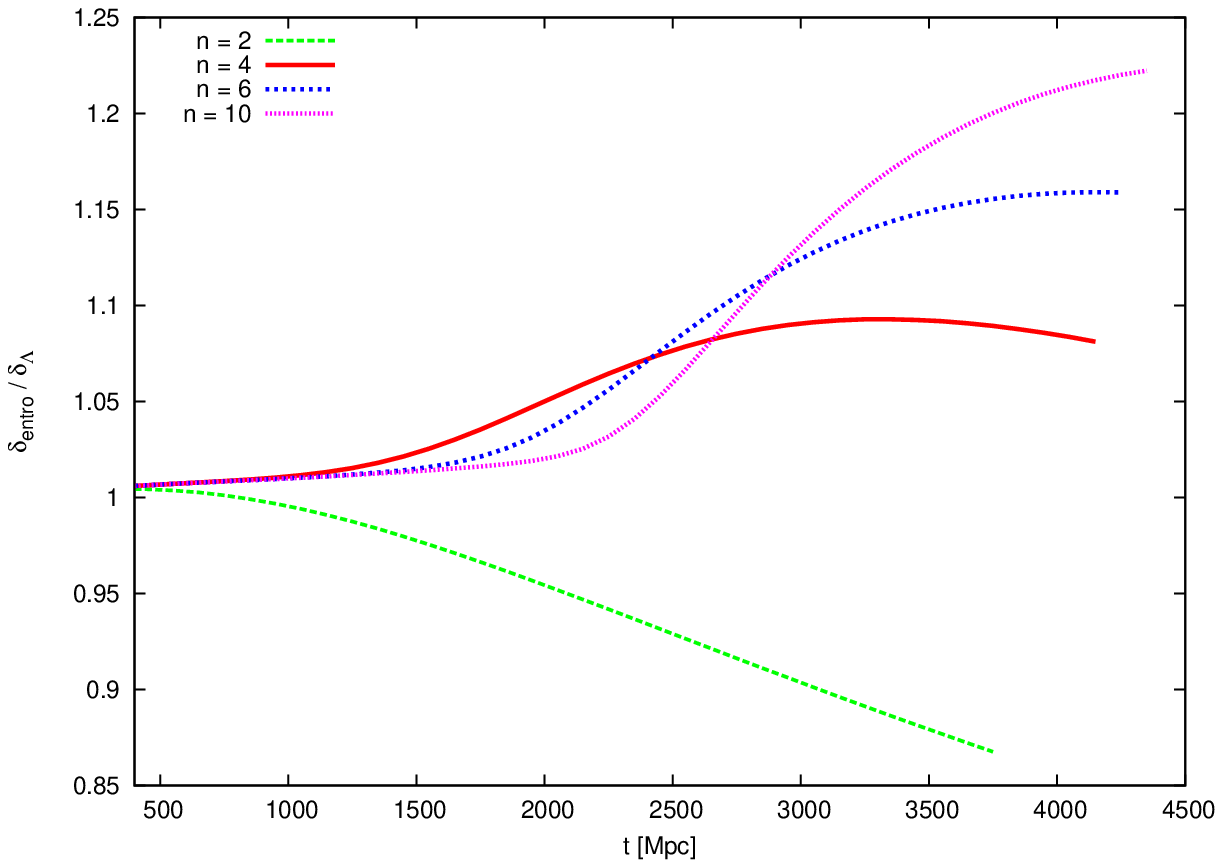}
\includegraphics[width=.49 \textwidth]{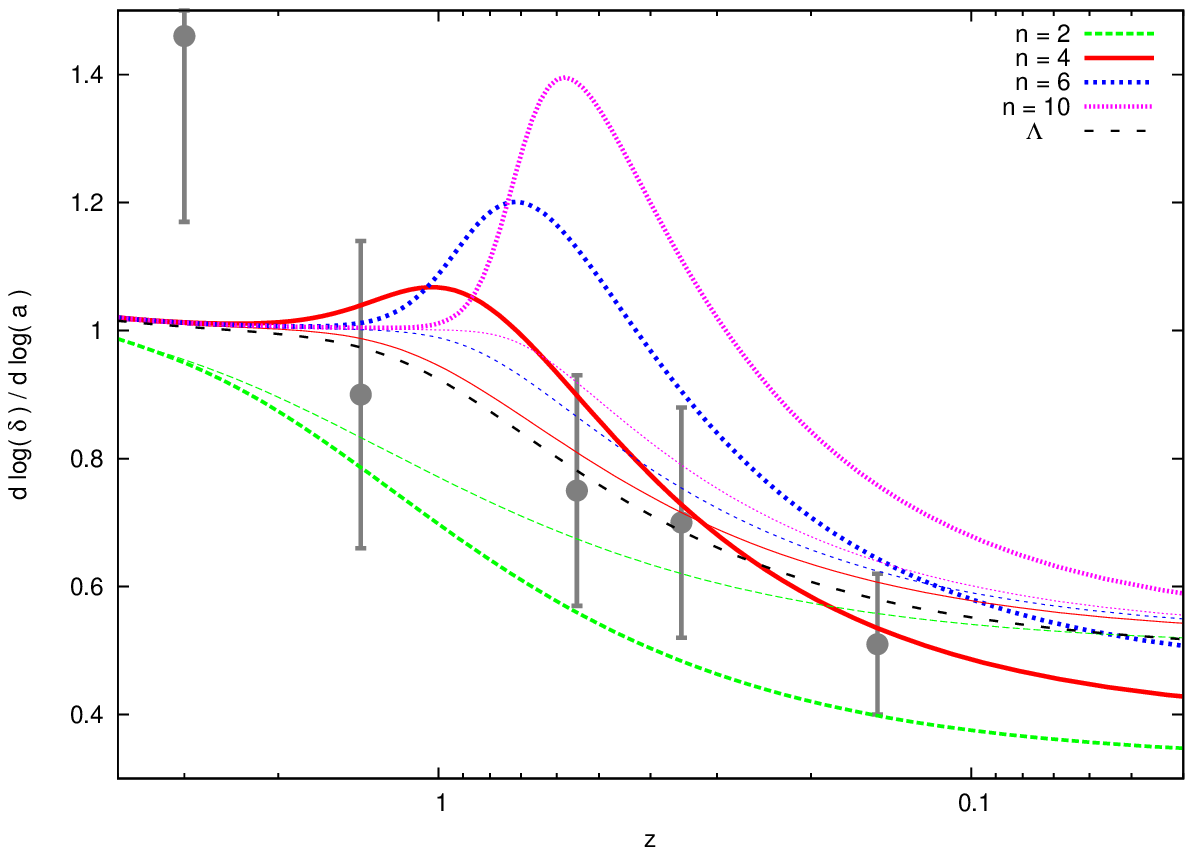}
\includegraphics[width=.49 \textwidth]{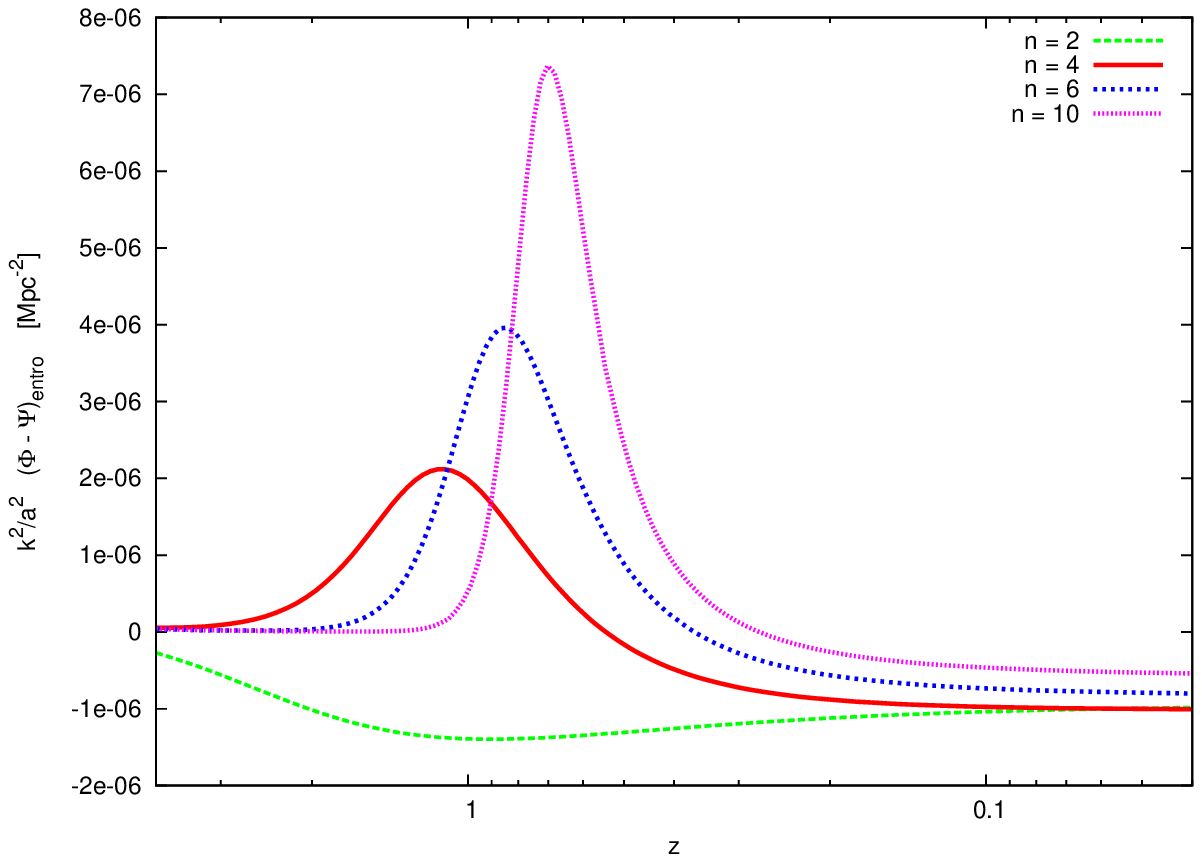}
\caption{Growth of structure in entropic cosmologies. \textbf{Top left:} Coefficient of $\delta$ in the r.h.s of (\ref{lue}) (thick lines) and (\ref{smooth}) (thin lines) for some values of $n$.
\textbf{Top right:} Time evolution of $\delta$ normalized to the $\Lambda$CDM value.
\textbf{Bottom left:} Evolution of the growth factor (\ref{growthfactor}) obtained using the modified equation (\ref{lue}) (thick lines) and the usual one (\ref{smooth}) (thin lines). Data points correspond to Table 1 in Ref. \cite{Nesseris:2007pa}.
\textbf{Bottom right:} Anisotropic stress caused by the modified gravitational potentials (\ref{lue2},\ref{lue3}). All lines correspond to flat universes with $\Omega_m=0.3$, $\Omega_S=0.7$.} \label{growth}
}

As an example we consider the power-law parametrization (\ref{powerlaw}). The growth rate $f$ of the perturbations can be defined as
\be\label{growthfactor}
f \equiv \frac{d\log{\delta}}{d\log{a}}\,.
\ee
 Asymptotically, when the the effective EoS is given by (\ref{weff}), the two solutions
to the evolution equation (\ref{lue}) correspond to growth rate $f=3/(2n)$ and $f=3/n-2$. The first one is the growing solution and thus
dominates at late times if $n>9/4$.
As $n$ increases, the decay of perturbations becomes less rapid, and when $n\rightarrow \infty$,
the background is de Sitter and the matter density is frozen. In dark energy cosmologies described by (\ref{smooth}), the solution with growing $\delta$ is absent and the transition behaviour, which is relevant to observations, is different.
The growing solution corresponds asymptotically to $\Phi,\Psi \sim a^p$, where $p=2-9/(2n)$ if $0<n<9/4$ and $p=0$ if $n>9/4$. Thus we expect that the gravitational potentials will tend asymptotically to a constant value if the present acceleration stems from entropic nature of gravity. This feature may produce the observational signatures of these models which could be used to distinguish them from dark energy models with the same background expansion. The growing solution to Eq.(\ref{smooth}) corresponds to $f=3/n-2$ and $p=-3/(2 n)$ when $n>3/4$. Thus in the case of smooth dark energy field, the gravitational potentials as well as the overdensities decay faster and vanish asymptotically. The relative amplification of the of gravitational potentials we find in the entropic cosmologies could be detected for instance in negative LSS-ISW correlations. Of course the present universe is only approaching the exact solutions mentioned above.

\TABLE{
\hspace{2.5cm} 
\begin{tabular}{l l  l l  }
 $n$  &  $a=1$ & $a=50$  &  $a\to\infty$ \\	\hline \\
 2 & 0.326 & -0.374 & -0.5  \\[3pt]
 4 & 0.390  & -0.359 & -0.375  \\[3pt]
 6 & 0.460  & -0.246 & -0.25  \\[3pt]
 10 & 0.533  & -0.149 & -0.15  \\[3pt]
\hline
\end{tabular}
\hspace{2.5cm} 
\caption{Numerical and asymptotic values of the growth factor $f$. \label{growthtable}}
}

Numerical solutions for the full evolution of perturbations are plotted in Figure \ref{growth}. Table \ref{growthtable} shows the numerical values of the growth factor (\ref{growthfactor}) at $a=1$ and $50$, together with the limit previously described. Discrepancies occur for the cases with low $n$, which have less negative e.o.s. and need longer time to reach the limit.
The growth of inhomogeneities can be enhanced or damped depending on the value of $n$.
The weakening of gravity responsible for the acceleration is reflected on smaller scales in the lower value of the r.h.s. factor of (\ref{lue}) responsible for gravitational instability. However, that term can become larger than $4\pi G\rho_m$ around the beginning of the acceleration due to the variation of $H$ through its time derivatives\footnote{Ref. \cite{Lue:2003ky} argue that slower growth of inhomogeneities follow from the acceleration condition. However, they consider a Friedman equation which is locally $H^2\sim \rho^n$ which is inequivalent to (\ref{friedmann}).}.
This effect can easily overcome the weakening of gravity if $n$ is large enough (more pronounced transition), leading to higher values of $\delta$. The enhancement effect is also responsible for the resulting anisotropic stress due to the same factors appearing in (\ref{lue}) and in (\ref{lue3}).

\subsection{The visible universe and the Bekenstein bound}

Reference \cite{Frampton:2010xu} raised an interesting point concerning a possible violation of the Bekenstein
bound by the
entropy contained in the visible universe. As is well known, it can be reasoned that the entropy $S$
of a region cannot exceed the entropy $S_{BH}$ of a black hole of the mass $M$ contained in the region,
\be \label{sbh}
S \le S_{BH} =  \frac{4\pi R_S^2}{\ell_P^2}\,,
\ee
$R_S=2GM$ being the Schwarzchild radius.
When the considered region is the whole visible universe, it is natural to take the radius to be given by $d_*$, the
comoving distance to the surface of last scattering. In holographic cosmology, the entropy (neglecting
corrections for now) is then just
\be \label{svu}
S_{VU} = \frac{4\pi d_*^2}{\ell_P^2}\,.
\ee
Now the mass (can be the Tolman-Komar) contained within the sphere of the radius $d_*$ is
\be \label{mass}
M=\frac{4\pi}{3}d_*^3\rho_M\,.
\ee
As reference \cite{Frampton:2010xu} showed, it is quite convenient to write these results in terms of the CMB shift
parameter $\mathcal{R}$ \cite{Bond:1997wr}
\be \label{shift}
\mathcal{R} \equiv \sqrt{\Omega_M}H_0 d_*\,.
\ee
Using Eq.(\ref{mass}) and then Eq.(\ref{shift}) we obtain that
\be
R_S=2GM=\frac{8\pi G\rho_M}{3}d_*^3 = H_0^2\Omega_M d_*^3=\mathcal{R}^2 d_*\,.
\ee
This enables us to write immediately the ratio
\be
\frac{S_{VU}}{S_{BH}} = \mathcal{R}^{-4}\,.
\ee
The observational constraint on the shift parameter is $\mathcal{R} = 1.725 \pm 0.018$ \cite{Komatsu:2010fb}.
The fact that the visible universe respects the entropy bound is a consequence of it being confined within its
own Schwarzchild radius. This is a nontrivial consistency check that the entropic cosmology passes. One can also
check that this continues to hold when the quantum corrections are taken into account. For simplicity
restricting to the flat case, we obtain then
\be \label{ratio}
\frac{S_{VU}}{S_{BH}} = \frac{1}{\mathcal{R}^4}
\frac{n\Omega_M+(1-2n)\Omega_s(d_*/H_0)^{2(n-1)}}{n\Omega_M+(1-2n)\Omega_s(\mathcal{R}^2d_*/H_0)^{2(n-1)}}\,.
\ee
As $n$ increases, the ratio gets smaller and the bound is fulfilled with a wider margin, as can be seen in Figure \ref{rpic}.

\FIGURE{
\includegraphics[width=.6 \textwidth]{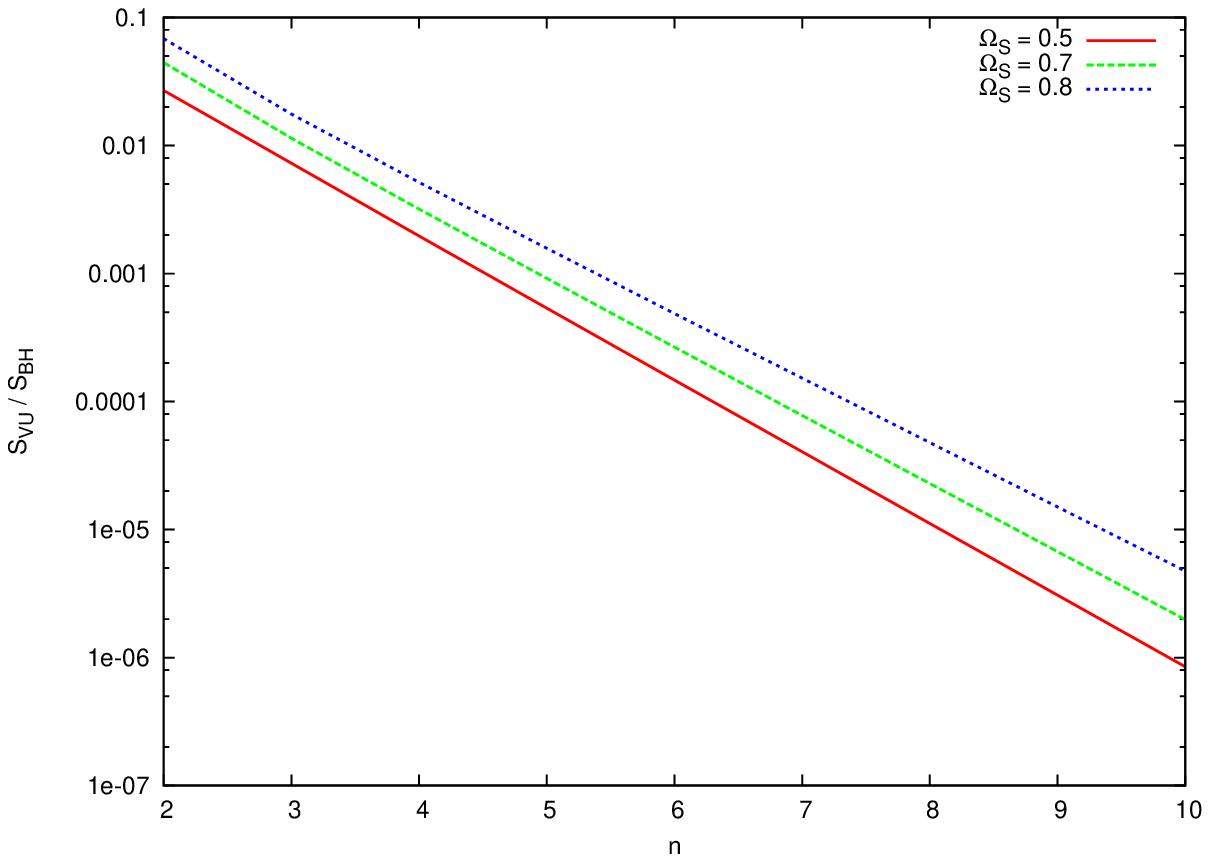}
\caption{The ratio of the entropy of the visible universe and of the black hole of the same mass in the model where the the monomial correction of power $n$ accelerates the universe. The different lines correspond to flat universes with different values of $\Omega_S$.}\label{rpic}
}

\subsection{The no-go theorem for inflation and its avoidance}

A further interesting observation is that these inflationary models may avoid the no-go theorem prohibiting inflation in entropic force law scenarios derived by reference \cite{Li:2010bc}. The theorem is based on the observation that the active gravitational mass becomes negative in accelerating cosmology, which implies negative temperature on the holographic screen. The active mass is considered to be the Tolman-Komar mass,
\be \label{tk}
M=2\int_\Sigma \lp T_{ab}-\frac{1}{2}T\rp n^a \xi^b dV\,,
\ee
where $\Sigma$ encloses the considered volume, $n^a$ is normal to it and $\xi^b$ is a time-like Killing vector. In the cases considered here, the Einstein equations are not satisfied, and thus $M$ can be positive even if the universe accelerates. Equating the two vectors with the four-velocity of matter, one obtains in FRW
\be \label{tk_frw}
M = \frac{4\pi}{3}(\rho+3 p)a^3 r^3\,.
\ee
In our examples we have radiation dominated universe where inflation is driven by entropic corrections, and clearly the Tolman-Komar mass (\ref{tk_frw}) is positive.

\section{Conclusions}
\label{conclusions}

In this paper we performed some elementary considerations of the possible consequences of modifications to the Friedman equations that have been suggested to describe the effects of holographic entropy on cosmology.

We found that the higher order curvature corrections, motivated by quantum corrections, lead to a graceful exit problem and thus can be, at least in the simplest scenarios, excluded. It was also observed that in two quite different approaches, the entropic corrections lead to a similar expansion law that predicts inflation in the early universe. This inflationary period has a natural transition to a radiation dominated universe. In the surface term approach of section \ref{surface}, we identified the parameter combinations (\ref{c_i}) that can be constrained by observations.
There $c_1$ quantifies the lower order and $c_2$ the higher order contributions. We obtained:
\ba
&-17.28 \cdot 10^{-3}  <  2c_1-3 < 20.50 \cdot 10^{-3}\,,& \nonumber \\
& \sqrt{|8\pi G c_2|}  <  0.02 \quad 1/GeV \,. &  \nonumber
\ea
from late and early universe constraints, respectively. In an alternative prescription retaining the cosmological matter conservation laws, previously introduced and bounded \cite{Casadio:2010fs}, an estimate for the parameters can be given as
\be
\bt_1-\al_1 = 0.02 \pm 0.08 \,.
\ee
In the quantum corrected approach discussed in section \ref{loop}, the leading logarithm correction to the entropy-area relation was shown to be constrained by BBN in a similar way, and the following inverse correction was the cause of the graceful exit problem. 


In addition, we studied a phenomenological power-law correction to the entropy formula. It was found that such can generate accelerating expansion in the late universe. Combining the available data to bound on the power of the correction, we obtained
\be
n = 3.8 \pm 0.7 \,.
\ee
We may go significantly further if the evolution of spherical metrics can be argued to depend only on the amount of enclosed matter. Then the features of linearized structure is encaptured by the three equations (\ref{lue},\ref{lue2},\ref{lue3}). 

We also pointed out that the inflation realized by the correction terms may avoid the no-go theorem prohibiting inflation driven by material sources in entropic cosmologies. In this light, we may claim to have not only have verified that the entropic corrections {\it can} drive inflation and the present acceleration of the universe but that they {\it must} be responsible for it, if the entropic emergence proposal in reference \cite{Verlinde:2010hp} is true. Coupled with the hints in reference \cite{Modesto:2010rm} that quantum corrections to entropy may also eliminate the need for dark matter, this would suggest a drastic reinterpretation of cosmological observations within an entropic paradigm.

\section*{acknowledgements}

TK is supported by the Academy of Finland
and the Yggdrasil grant of the Research Council of Norway.  DFM thanks the Research Council of Norway FRINAT grant 197251/V30
and the Abel extraordinary chair UCM-EEA-ABEL-03-2010. DFM is also
partially supported by the projects CERN/FP/109381/2009 and
PTDC/FIS/102742/2008. MZ is funded by MICINN (Spain) through the project AYA2006-05369 and the grant BES-2008-009090.


\bibliography{hrefs}





\end{document}